\documentclass[aps,pra,twocolumn,nobibnotes,superscriptaddress,nofootinbib,10pt]{revtex4-1} 
\usepackage[latin1]{inputenc}
\usepackage{amsmath,amssymb}
\usepackage{mathrsfs} 
\usepackage{graphicx,color,colortbl}
\usepackage{rotating,array,tabularx,booktabs}
\usepackage{varwidth,xcolor}
\usepackage{placeins}
\usepackage{siunitx}

\DeclareGraphicsExtensions{.pdf,.PDF,.png,.PNG}

\newcommand{\dif}{\mathrm{d}}%
\newcommand{\uu}{\hat{n}}%
\newcommand{\ww}{\vec{u}}%
\newcommand{\ws}{u}%

\begin{document}

\title{Orientation-dependent propulsion of cone-shaped nano- and microparticles by a traveling ultrasound wave}

\author{Johannes Vo\ss{}}
\affiliation{Institut f\"ur Theoretische Physik, Center for Soft Nanoscience, Westf\"alische Wilhelms-Universit\"at M\"unster, D-48149 M\"unster, Germany}

\author{Raphael Wittkowski}
\email[Corresponding author: ]{raphael.wittkowski@uni-muenster.de}
\affiliation{Institut f\"ur Theoretische Physik, Center for Soft Nanoscience, Westf\"alische Wilhelms-Universit\"at M\"unster, D-48149 M\"unster, Germany}

\begin{abstract}
Previous studies on ultrasound-propelled nano- and microparticles have considered only systems where the particle orientation is perpendicular to the direction of propagation of the ultrasound. However, in future applications of these particles, they will typically be able to attain also other orientations. Therefore, using direct acoustofluidic simulations, we here study how the propulsion of cone-shaped nano- and microparticles, which are known to have a particularly efficient acoustic propulsion and are therefore promising candidates for future applications, depends on their orientation relative to the propagation direction of a traveling ultrasound wave. Our results reveal that the propulsion of the particles depends strongly on their orientation relative to the direction of wave propagation and that the particles tend to orient perpendicularly to the wave direction. We also present the orientation-averaged translational and angular velocities of the particles, which correspond to the particles' effective propulsion for an isotropic exposure to ultrasound. Our results allow assessing how free ultrasound-propelled colloidal particles move in three spatial dimensions and thus constitute an important step towards the realization of the envisaged future applications of such particles. 
\begin{figure}[htb]
\centering
\fbox{\includegraphics[width=8cm]{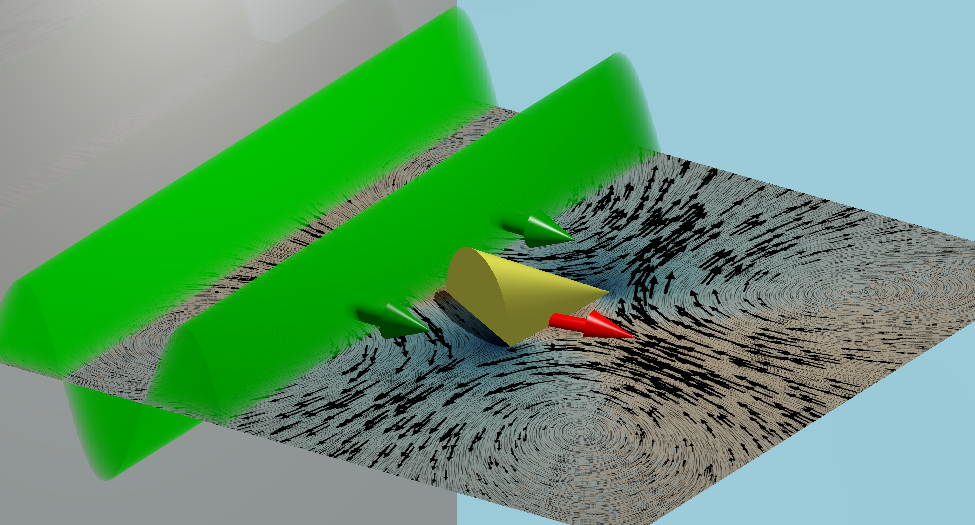}}%
\end{figure}
\end{abstract}
\maketitle

\section{Introduction}
The experimental discovery of colloidal particles with fuel-free ultrasound propulsion in 2012 \cite{WangCHM2012} made important applications of active particles \cite{BechingerdLLRVV2016} become within reach \cite{XuXZ2017,Venugopalan2020,FernandezRMHSS2020,YangEtAl2020}. Especially in medicine \cite{LiEFdAGZW2017,PengTW2017,SotoC2018,WangGZLH2020,WangZ2021}, where such particles could be used, e.g., for targeted drug delivery \cite{LuoFWG2018,ErkocYCYAS2019}, and in materials science \cite{WangDZLGYLWM2019}, where they could form active materials with exceptional properties \cite{JunH2010,McdermottKDKWSV2012}, ultrasound-propelled nano- and microparticles \cite{WangCHM2012,GarciaGradillaEtAl2013,AhmedEtAl2013,WuEtAl2014,WangLMAHM2014,GarciaGradillaSSKYWGW2014,BalkEtAl2014,AhmedGFM2014,EstebanFernandezdeAvilaMSLRCVMGZW2015,WuEtAl2015a,WuEtAl2015b,RaoLMZCW2015,EstebanEtAl2016,SotoWGGGLKACW2016,AhmedWBGHM2016,AhmedBJPDN2016,UygunEtAl2017,KaynakONLCH2017,EstebanFernandezEtAl2017,HansenEtAl2018,SabrinaTABdlCMB2018,WangGWSGXH2018,EstebanEtAl2018,LuSZWPL2019,QualliotineEtAl2019,GaoLWWXH2019,RenEtAl2019,VossW2020,VossW2021,AghakhaniYWS2020,LiuR2020,ZhouYWDW2017,ZhouZWW2017,ValdezLOESSWG2020,DumyJMBGMHA2020} have great potential. Since acoustic propulsion is biocompatible and allows to supply the particles permanently with energy, ultrasound-propelled particles are more suitable for medical applications than other propulsion mechanisms that have been developed in the past \cite{EstebanFernandezdeAvilaALGZW2018,SafdarKJ2018,PengTW2017,KaganBCCEEW2012,XuanSGWDH2018,XuCLFPLK2019}. 

The great potential for applications of ultrasound-propelled nano- and microparticles resulted in an intensive investigation of their properties \cite{WangCHM2012,GarciaGradillaEtAl2013,AhmedEtAl2013,NadalL2014,WuEtAl2014,WangLMAHM2014,GarciaGradillaSSKYWGW2014,BalkEtAl2014,AhmedGFM2014,AhmedLNLSMCH2015,WangDZSSM2015,EstebanFernandezdeAvilaMSLRCVMGZW2015,WuEtAl2015a,WuEtAl2015b,RaoLMZCW2015,KimGLZF2016,EstebanEtAl2016,SotoWGGGLKACW2016,AhmedWBGHM2016,KaynakONNLCH2016,UygunEtAl2017,EstebanFernandezEtAl2017,RenZMXHM2017,KaynakONLCH2017,CollisCS2017,ZhouZWW2017,ZhouYWDW2017,ChenEtAl2018,RenWM2018,HansenEtAl2018,SabrinaTABdlCMB2018,AhmedBJPDN2016,Zhou2018,WangGWSGXH2018,EstebanEtAl2018,LuSZWPL2019,TangEtAl2019,QualliotineEtAl2019,GaoLWWXH2019,RenEtAl2019,AghakhaniYWS2020,LiuR2020,VossW2020,ValdezLOESSWG2020,DumyJMBGMHA2020,VossW2021}.  
Most of the existing studies are experimental \cite{WangCHM2012,GarciaGradillaEtAl2013,AhmedEtAl2013,WuEtAl2014,WangLMAHM2014,GarciaGradillaSSKYWGW2014,BalkEtAl2014,AhmedGFM2014,AhmedLNLSMCH2015,WangDZSSM2015,EstebanFernandezdeAvilaMSLRCVMGZW2015,WuEtAl2015a,WuEtAl2015b,EstebanEtAl2016,SotoWGGGLKACW2016,AhmedWBGHM2016,KaynakONNLCH2016,UygunEtAl2017,KaynakONLCH2017,EstebanFernandezEtAl2017,RenZMXHM2017,ZhouYWDW2017,HansenEtAl2018,RenWM2018,SabrinaTABdlCMB2018,AhmedBJPDN2016,ZhouZWW2017,WangGWSGXH2018,EstebanEtAl2018,Zhou2018,TangEtAl2019,QualliotineEtAl2019,GaoLWWXH2019,RenEtAl2019,AghakhaniYWS2020,LiuR2020,ValdezLOESSWG2020,DumyJMBGMHA2020}, a few of them are numerical \cite{AhmedBJPDN2016,SabrinaTABdlCMB2018,Zhou2018,TangEtAl2019,VossW2020,VossW2021}, and only two studies are analytical \cite{NadalL2014,CollisCS2017}. 
The previous studies considered both rigid particles \cite{WangCHM2012,GarciaGradillaEtAl2013,AhmedEtAl2013,NadalL2014,BalkEtAl2014,AhmedGFM2014,GarciaGradillaSSKYWGW2014,WangLMAHM2014,WangDZSSM2015,EstebanFernandezdeAvilaMSLRCVMGZW2015,SotoWGGGLKACW2016,AhmedWBGHM2016,EstebanEtAl2016,KaynakONNLCH2016,UygunEtAl2017,CollisCS2017,HansenEtAl2018,EstebanEtAl2018,SabrinaTABdlCMB2018,Zhou2018,RenWM2018,TangEtAl2019,ZhouZWW2017,VossW2020,VossW2021,ValdezLOESSWG2020,DumyJMBGMHA2020} and particles with movable components \cite{KaganBCCEEW2012,AhmedLNLSMCH2015,AhmedBJPDN2016,KaynakONLCH2017,WangGWSGXH2018,RenEtAl2019,AghakhaniYWS2020,LiuR2020}. 
In the former case, various particle shapes have been considered: cylinders with concave and convex ends \cite{WangCHM2012,GarciaGradillaEtAl2013,AhmedEtAl2013,GarciaGradillaSSKYWGW2014,AhmedGFM2014,BalkEtAl2014,WangLMAHM2014,EstebanFernandezdeAvilaMSLRCVMGZW2015,EstebanEtAl2016,AhmedWBGHM2016,UygunEtAl2017,ZhouZWW2017,CollisCS2017,EstebanEtAl2018,HansenEtAl2018,Zhou2018,DumyJMBGMHA2020}, half-spheres \cite{VossW2020}, half-sphere cups (nanoshells) \cite{SotoWGGGLKACW2016,TangEtAl2019, VossW2020}, cones \cite{VossW2020,VossW2021}, and gear-shaped particles \cite{KaynakONNLCH2016,SabrinaTABdlCMB2018}. 
Among the studied particles, cone-shaped ones have been shown to be particularly promising candidates for future research and applications, since they can easily be produced in large numbers and have relatively efficient propulsion \cite{VossW2021}. 
Even hybrid particles combining acoustic propulsion with other propulsion mechanisms have been studied \cite{LILXKLWW2015,WangDZSSM2015,RenZMXHM2017,TangEtAl2019,Zhou2018,RenWM2018,ValdezLOESSWG2020}. 
Nevertheless, our understanding of ultrasound-propelled nano- and microparticles is still very incomplete. 

One problem is that nearly all existing studies consider a standing ultrasound wave, although for future applications a traveling ultrasound wave is much more realistic \cite{VossW2020,VossW2021}.  
Another problem is that in the previous studies the particles are oriented perpendicular to the propagation direction of the ultrasound, although we can expect that in future applications the particles will be able to orient also differently when they move, e.g., within an active suspension forming an active material or within the vascular system of a patient. 
The reason for studying only perpendicular orientations so far is that in the experimental studies the particles are levitated in the nodal plane of a standing ultrasound wave, which is perpendicular to the propagation direction of the ultrasound wave and restricts the motion and orientation of the particles to that plane, and that the existing theoretical studies consider systems that are similar to those that have been investigated in experiments. 

In this work, we go an important step further by studying the acoustic propulsion of particles that are exposed to a planar traveling ultrasound wave and that can orient in any direction relative to the ultrasound wave. 
Focusing on the promising cone-shaped particles, we study how their propulsion depends on the relative orientation of particle and ultrasound wave. 
For this purpose, we apply direct computational fluid dynamic simulations that are based on the compressible Navier-Stokes equations to calculate the propagation of the ultrasound and its interaction with a particle. These simulations yield the sound-induced forces and torques acting on the particle, which in turn determine its translational and angular propulsion velocity.

\section{\label{results}Results and discussion}
We determined the time-averaged propulsion of a cone-shaped particle in water that is exposed to a planar traveling ultrasound wave with frequency $f=\SI{1}{\MHz}$ and energy density $E=\SI{22.7}{\milli\joule\,\metre^{-3}}$ for various orientations of the particle.
The considered particle has diameter $\sigma=2^{-1/2}\,\SI{}{\micro\metre}$ and height $h=\sigma$ and the water is initially at standard temperature, under standard pressure, and quiescent. See Methods for details. 

\begin{figure}[htb!]
\centering
\includegraphics[width=\linewidth]{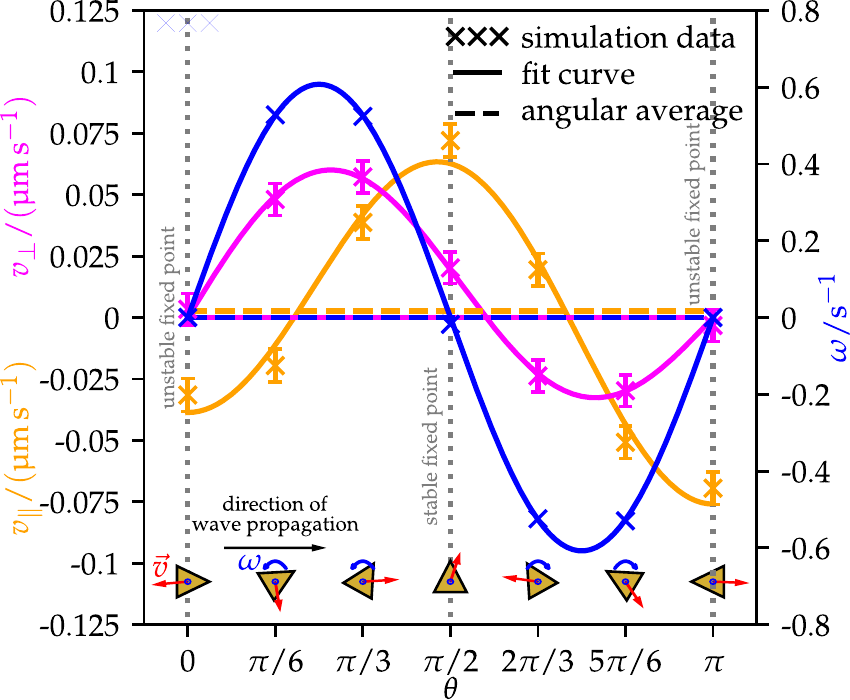}%
\caption{\label{fig:fig1}Simulation data and fit curves for the translational velocity components $v_\parallel$ and $v_\perp$ parallel and perpendicular to the orientation of a cone-shaped particle, respectively, as well as the particle's angular velocity $\omega$ for various orientations of the particle. In addition, the orientationally averaged values of $v_\parallel$, $v_\perp$, and $\omega$, as well as fixed points of the particle orientation, are shown. See Ref.\ \cite{VossW2021bSI} for the raw data corresponding to this figure.}
\end{figure}

Figure \ref{fig:fig1} shows our results for the particle's translational propulsion velocity $v_\parallel$ parallel to the orientation (i.e., symmetry axis) of the particle, the component $v_\perp$ perpendicular to the particle's orientation, and the angular velocity $\omega$ relative to the particle's center of mass for orientations $\theta\in[0,\pi]$. At $\theta=0$, the particle orientation and the propagation direction of the ultrasound are parallel, and at $\theta=\pi$, they are antiparallel.   
Remarkably, the considered ultrasound-propelled particle has orientation-dependent propulsion.

The velocity component $v_\parallel$ starts with a local minimum $v_\parallel=\SI{-0.032\pm0.007}{\micro\metre\,\second^{-1}}$ at $\theta=0$. This means that for a particle that is oriented parallel to the ultrasound wave, the velocity component $v_\parallel$ moves the particle backwards and thus antiparallel to the ultrasound. 
$v_\parallel$ then increases, changes its sign between $\theta=\pi/6$ and $\theta=\pi/3$, and further increases until it reaches its maximum $v_\parallel=\SI{0.072\pm 0.007}{\micro\metre\,\second^{-1}}$ at $\theta=\pi/2$ where particle and ultrasound are oriented perpendicularly. Hence, in the situation of all the previous studies on ultrasound-propelled particles, where the particles are perpendicular to the ultrasound, the particles considered in the present work exhibit their fastest forward motion. 
For even larger values of the angle $\theta$, $v_\parallel$ decreases again. Between $\theta=2\pi/3$ and $\theta=5\pi/6$ it changes sign for the second time and it reaches its global minimum $v_\parallel=\SI{-0.069\pm0.007}{\micro\metre\,\second^{-1}}$ at $\theta=\pi$. Here, the particle again moves backward, which now means parallel to the ultrasound.     
To find a simple function that interpolates the simulation data for $v_\parallel$, we take into account that it must have the symmetry property $v_\parallel(\theta) = v_\parallel(-\theta)$ to reflect the symmetry of the particle shape. We, therefore, use the second-order Fourier series 
\begin{equation}
v_\parallel(\theta) = a_{\parallel,0} + a_{\parallel,1}\cos(\theta) + a_{\parallel,2} \cos(2\theta), 
\label{eq:vparallelfit}%
\end{equation}
which is in good agreement with the simulation results.
The values of the expansion coefficients $a_{\parallel,0}$, $a_{\parallel,1}$, and $a_{\parallel,2}$ that result from fitting the function \eqref{eq:vparallelfit} to the experimental data are given in Tab.\ \ref{tab:Fitparameter}.  
\begin{table}[tb]
\centering
\begin{ruledtabular}
\begin{tabular}{@{}cccc@{}}
$\boldsymbol{\zeta}$ & $\mathbf{\boldsymbol{a}_{\boldsymbol{\zeta},0}}$ & $\mathbf{\boldsymbol{a}_{\boldsymbol{\zeta},1}}$ & $\mathbf{\boldsymbol{a}_{\boldsymbol{\zeta},2}}$\\
\hline
$\parallel$ & $ \SI{0.002691}{\micro\metre\,\second^{-1}}$ & $\SI{0.01864}{\micro\metre\,\second^{-1}} $ &  $\SI{-0.05999}{\micro\metre\,\second^{-1}}$ \\
$\perp$ & $-$ & $\SI{0.01949}{\micro\metre\,\second^{-1}}$ & $ \SI{0.09169}{\micro\metre\,\second^{-1}}$ 
\end{tabular} 
\end{ruledtabular}%
\caption{\label{tab:Fitparameter}Fit parameters for the velocity components $v_\zeta$ with $\zeta\in\{\parallel,\perp\}$. 
See Fig.\ \ref{fig:fig3} for their definitions.}
\end{table}
\begin{figure*}[tb]
\centering
\includegraphics[width=\linewidth]{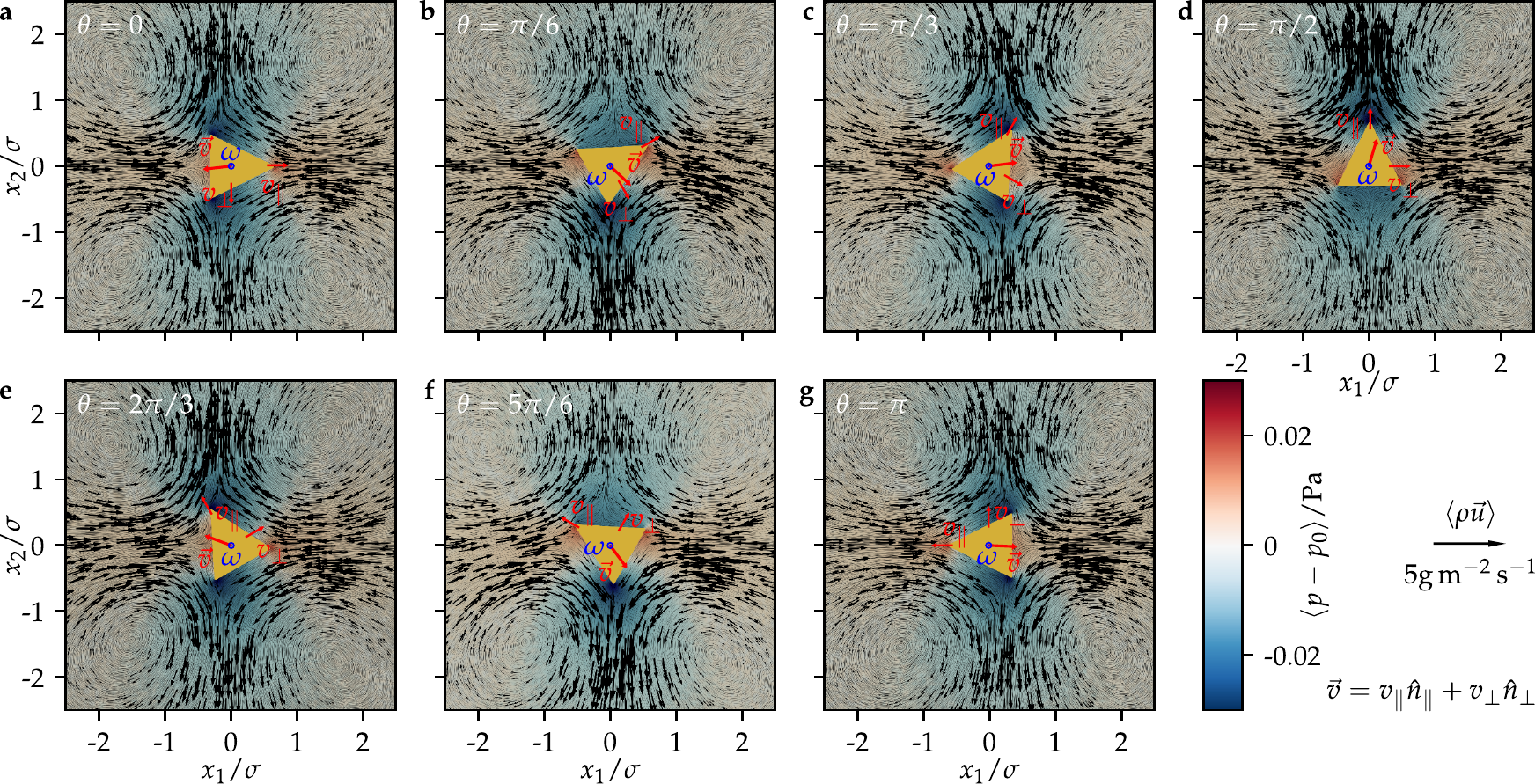}%
\caption{\label{fig:fig2}The time-averaged mass-current density $\langle\rho\ww\rangle$ and reduced pressure $\langle p-p_{0}\rangle$ as well as the translational propulsion velocity $\vec{v}=v_\parallel \uu_\parallel + v_\perp \uu_\perp$ and angular propulsion velocity $\omega$ for all considered orientations of the particle, where $\rho$, $\ww$, and $p$ are the fluid's mass density field, velocity field, and pressure field, respectively, and $\uu_\parallel$ and $\uu_\perp$ are orientation vectors that are parallel and perpendicular to the particle orientation, respectively. 
See Ref.\ \cite{VossW2021bSI} for the raw data corresponding to this figure.}
\end{figure*}
An angular average of $v_\parallel$ yields
\begin{equation}
\langle v_\parallel\rangle_\theta 
= \frac{1}{2\pi} \int_0^{2\pi}\!\!\!\!\!\!\! v_\parallel(\theta)  \,\mathrm{d}\theta 
=a_{\parallel,0}
=\SI{0.00269}{\micro\metre\,\second^{-1}} . 
\label{eq:Spannungstensorv}%
\end{equation}
This means that for an isotropic ultrasound field, the particle will move forward. 

For the velocity component $v_\perp$, qualitatively different behavior is observed. 
This component vanishes at $\theta=0$ for reasons of symmetry, meaning that there is no motion perpendicular to the ultrasound wave when the particle is oriented parallel to the propagation direction of the ultrasound. 
When $\theta$ increases, the value of $v_\perp$ increases until a maximum is reached between $\theta=\pi/6$, where $v_\perp=\SI{0.048\pm0.007}{\micro\metre\,\second^{-1}}$, and $\theta=\pi/3$, where $v_\perp=\SI{0.057\pm0.007}{\micro\metre\,\second^{-1}}$.  
Afterwards, $v_\perp$ decreases, changes its sign between $\theta=\pi/2$ and $\theta=2\pi/3$, and reaches its minimum between $\theta=2\pi/3$, where $v_\perp=\SI{-0.024\pm 0.007}{\micro\metre\,\second^{-1}}$, and $\theta=5\pi/6$, where $v_\perp=\SI{-0.03\pm 0.007}{\micro\metre\,\second^{-1}}$. 
For larger angles $\theta$, the component $v_\perp$ increases again until it vanishes at $\theta=\pi$ for reasons of symmetry.  
This means that when the particle's orientation has a component parallel to the propagation direction of the ultrasound, the velocity component $v_\perp$ contributes to a parallel motion of the particle, whereas for an orientation with a moderate or large antiparallel component $v_\perp$ contributes to an antiparallel motion.
Note that due to numerical inaccuracies in the calculations, the values of $v_\perp$ are not exactly zero at $\theta=0$ and $\theta=\pi$.
For interpolating the simulation results for $v_\perp$, we take into account the symmetry property $v_\perp(\theta)=-v_\perp(-\theta)$ that results from the setup of the considered system and use the simple function  
\begin{equation}
v_\perp(\theta) = a_{\perp,1}\sin(\theta) + a_{\perp,2}\sin(\theta)\cos(\theta)
\end{equation}
that is in good agreement with the simulation data. 
The values of the expansion coefficients $a_{\perp,1}$ and $a_{\perp,2}$ that result from fitting this function to the simulation data are given in Tab.\ \ref{tab:Fitparameter}.  
For reasons of symmetry, the orientation-averaged velocity $\langle v_\perp\rangle_\theta$ vanishes. 

For the angular velocity $\omega$, we find a qualitatively similar behavior as for $v_\perp$, but $\omega$ vanishes also for $\theta=\pi/2$, although this follows not directly from symmetry considerations.
The particle therefore rotates anticlockwise for angles $0<\theta<\pi/2$ and clockwise for $\pi/2<\theta<\pi$.  
This means that there are unstable fixed points of the particle orientation at $\theta=0$ and $\theta=\pi$ and a stable fixed point at $\theta=\pi/2$.
The particle thus tends to orient perpendicular to the direction of ultrasound propagation. 
It is interesting that a perpendicular alignment of the particle, which has been observed in experiments with standing ultrasound waves \cite{WangCHM2012,GarciaGradillaEtAl2013,BalkEtAl2014,SotoWGGGLKACW2016,AhmedWBGHM2016,ZhouZWW2017,SabrinaTABdlCMB2018,TangEtAl2019,DumyJMBGMHA2020}, occurs also here, where a traveling ultrasound wave is chosen.  
Using the rotational diffusion coefficient $D_\mathrm{R}=\SI{1.78}{\second^{-1}}$ of the particle that is considered in the present work, which is equivalent to a reorientation of the particle within $\SI{0.56}{\second}$, and comparing it with the maximal observed angular velocity of $\omega=\SI{0.53}{\second^{-1}}$, which implies a rotation by $\pi/2$ within about $\SI{1}{\second}$, we find that the reorientation of a particle by rotational propulsion is of the same order of magnitude as the reorientation by Brownian motion. Note that this applies to the energy density of the ultrasound that is chosen in this work. For lower or higher ultrasound intensities, Brownian motion or active rotation dominate, respectively. 
To interpolate the simulation data for $\omega$, we use the function 
\begin{equation}
\omega(\theta) = \SI{0.6079}{}\sin(2\theta)\,\SI{}{\second^{-1}} ,
\end{equation}
where the value of the prefactor is determined by fitting the function to the simulation data. 
Again, the agreement with the simulation data is good. 
The orientational average $\langle \omega \rangle_\theta$ of the angular velocity vanishes for reasons of symmetry.  

Assuming that the propulsion velocities of the particle are proportional to the energy density of the ultrasound, which is suggested by the fact that both the translational propulsion velocity \cite{AhmedBJPDN2016} and the energy density of the ultrasound \cite{Bruus2012} have been found to be proportional to the square of the driving voltage, we can estimate the values of the propulsion velocities for a higher ultrasound intensity. 
Increasing the energy density from $E=\SI{22.7}{\milli\joule\,\metre^{-3}}$, which corresponds to our simulations, to $E_\mathrm{max}=\SI{4.9}{\joule\,\metre^{-3}}$, which is the maximal energy density that is allowed by the U.S.\ Food and Drug Administration for diagnostic applications of ultrasound in the human body \cite{QBarnettEtAl2000}, can then be expected to lead to an increase of $v_\parallel$, $v_\perp$, and $\omega$ by a factor of about $216$. The orientationally averaged propulsion should then equal a forward propulsion speed $\langle v_{\parallel}\rangle_{\theta,\mathrm{rescaled}}\approx\SI{0.6}{\micro\metre\,\second^{-1}}$.  
Finally, we consider the flow field around the particle for different orientations. This is important to clarify whether the flow field around such an ultrasound-propelled particle is similar to that of a squirmer, which has been suggested by a recent study \cite{VossW2020}. This study found a squirmer-like flow field for different particles but considered only particle orientations perpendicular to the direction of ultrasound propagation. 
Our results for the flow field are shown in Fig.\ \ref{fig:fig2}. 
Interestingly, the flow field looks qualitatively similar irrespective of the orientation of the particle. 
This reveals that these particles cannot be considered as squirmers.

\section{\label{conclusions}Conclusions}
Our investigation of the motion of a cone-shaped particle that is propelled by a planar traveling ultrasound wave and has a variable orientation relative to the propagation direction of the ultrasound resulted in several important observations.

First, we found that the propulsion of the particle depends on its orientation. This is a feature that affects the dynamics of the particle in an interesting way, as has recently been addressed using particles with a different propulsion mechanism \cite{SprengerFRAIWL2020}.

Second, we revealed the particular orientation-dependence of the propulsion and provided simple analytical expressions for it. Knowledge of this dependence is very important with respect to future applications of acoustically propelled particles, e.g., in nanomedicine or materials science, where the particles will be able to attain various or all directions relative to the (typically traveling) ultrasound wave.
Compared to the previous studies from the literature that consider only particles that are perpendicular to a (standing) ultrasound wave, the present work thus constitutes a large step forward. 
Furthermore, the provided functions for the orientation-dependence of the propulsion can be used to model this propulsion when describing the behavior of the particles via Langevin equations \cite{WittkowskiL2012,tenHagenWTKBL2015} or field theories \cite{BickmannW2020twoD,BickmannW2020b,teVrugtLW2020}. For the future investigation of ultrasound-propelled nano- and microparticles, such a description would be highly advantageous, since its characteristic time scale can be orders of magnitude larger than the period of the ultrasound, which would strongly reduce the effort to study the particles' dynamics on times scales of seconds to hours as they correspond to experiments with such particles.

Third, we observed that, depending on the orientation of the particle, the velocity vector can show in any direction including an orientation antiparallel to the wave propagation. Also this finding is highly important with respect to applications, since it shows that the ultrasound-propelled particles can move even towards the source of the ultrasound.  

Fourth, we found that the orientation of the particle has three fixed points including two unstable ones and a stable one, where the latter fixed point corresponds to a particle orientation that is perpendicular to the ultrasound wave. This shows that the observation of previous studies \cite{WangCHM2012,GarciaGradillaEtAl2013,BalkEtAl2014,SotoWGGGLKACW2016,AhmedWBGHM2016,ZhouZWW2017,SabrinaTABdlCMB2018,TangEtAl2019,DumyJMBGMHA2020}, that the particles align within the nodal plane of a standing ultrasound wave, is not simply a result of the particles' levitation in the nodal plane but, at least partially, a result of an alignment mechanism that is present also for traveling ultrasound waves. This alignment mechanism is interesting since it provides new opportunities to steer acoustically propelled particles. 

Fifth, our results show a nonzero orientation-averaged forward propulsion of the particle. In an isotropic ultrasound field, the particle will therefore move forward irrespective of its orientation. This is an important finding since it shows that acoustically-propelled particles can move also simply by forward translation without rotational propulsion, like the idealized active particles that are primarily studied in the literature  \cite{BechingerdLLRVV2016,WittkowskiSC2017,BickmannW2020twoD,BickmannW2020b,JeggleSW2020,BroekerSW2021}. Such ultrasound-propelled particles can therefore be applied also in situations where purely translational propulsion is required (e.g., when performing experiments that correspond to the aforementioned studies), when the ultrasound supply is designed accordingly.

Sixth, we observed that the flow field around the particle looks rather similar for all of its orientations. This means that the particle cannot be described as a pusher, as has been assumed previously \cite{VossW2020}. On the other hand, it seems that one can describe the flow field by a pusher-like flow field that translates with the particle but has a fixed orientation. Based on this model for the flow field, it should be possible to determine the locally time-averaged hydrodynamic interactions between different ultrasound-propelled particles when they are not too close together. 

In summary, this work solves problems and provides new insights that are highly important on the way towards the intriguing applications that have been envisaged for ultrasound-propelled nano- and microparticles \cite{WangDZLGYLWM2019,JunH2010,McdermottKDKWSV2012,LiEFdAGZW2017,PengTW2017,SotoC2018,WangGZLH2020,WangZ2021,LuoFWG2018,ErkocYCYAS2019}. 
In the future, this work should be continued by varying the particle shape, particle size, ultrasound frequency, ultrasound intensity, and other parameters of the system and studying how this affects, e.g., the orientation-averaged propulsion velocity or the fixed points of the particle orientation.

\section{\label{methods}Methods}
\begin{figure}[htb]
\centering
\includegraphics[width=\linewidth]{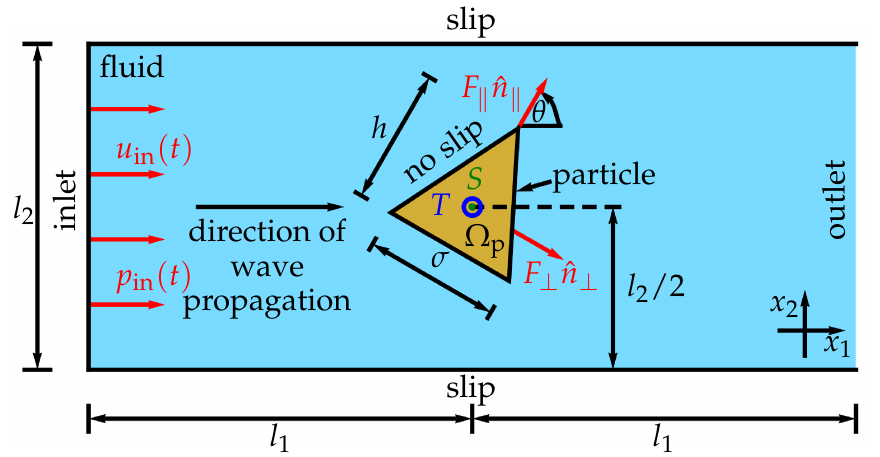}%
\caption{\label{fig:fig3}Setup for the simulations. The simulation domain has width (parallel to the $x_1$ axis of our coordinate system) $2l_1$ and height (parallel to the $x_2$ axis) $l_2$ and is filled with water as fluid. 
At the left boundary of the rectangular fluid domain, the inlet, a planar ultrasound wave enters the system and propagates in the $x_1$ direction.  
We prescribe the incoming ultrasound wave at this boundary by a time-dependent inflow velocity $\ws_{\mathrm{in}}(t)$ and inflow pressure $p_{\mathrm{in}}(t)$.
In the middle of the fluid domain, a rigid cone-shaped particle is placed so that the center of mass $S$ of the particle and that of the fluid domain coincide. This particle shape is chosen since it corresponds to a fast forward motion \cite{VossW2021}.
The particle constitutes a particle domain $\Omega_\mathrm{p}$. It has diameter $\sigma$, height $h$, and orientation angle $\theta$. We measure the angle $\theta$ in such a way that it describes the orientation of the particle, which is here defined as the orientation of the unit vector $\uu_{\parallel}$ that is parallel to the particle's symmetry axis and points from $S$ towards one tip of the particle, relative to the incoming ultrasound wave. For $\theta=0$, the particle's orientation $\uu_{\parallel}$ is parallel to the direction of wave propagation, for $\theta=\pi/2$ they are perpendicular, and for $\theta=\pi$, they are antiparallel. We define also an orientational unit vector $\uu_{\perp}$ that is always perpendicular to $\uu_{\parallel}$ and for $\theta=\pi/2$ parallel to the direction of wave propagation. 
The interaction of the ultrasound with the particle results in a time-averaged propulsion force with components $F_\parallel$ parallel to $\uu_{\parallel}$ and $F_\perp$ parallel to $\uu_{\perp}$ as well as in a time-averaged propulsion torque $T$ acting at $S$. 
At the right boundary of the fluid domain, the outlet, the ultrasound can leave the system. 
We use slip boundary conditions for the boundaries of the fluid domain that are parallel to the $x_1$ direction and no-slip boundary conditions for the boundary of the particle.}
\end{figure}

Figure \ref{fig:fig3} shows and explains the setup chosen for our simulations.
We use a rectangular simulation domain with width $2l_1$ and height $l_2=\SI{200}{\micro\metre}$ that is aligned with the coordinate system so that the width is along the $x_1$ axis and the height along the $x_2$ axis. The simulation domain is filled with water that initially is at standard temperature $T_0=\SI{293.15}{\kelvin}$ and standard pressure $p_0=\SI{101325}{\pascal}$ and has a vanishing velocity field $\ww_0=\vec{0}\,\SI{}{\metre\,\second^{-1}}$. In the middle of the simulation domain, a cone-shaped particle with diameter $\sigma=2^{-1/2}\,\SI{}{\micro\metre}$, height $h=\sigma$, and particle domain $\Omega_\mathrm{p}$ is placed such that the center of masses $S$ of simulation domain and particle coincide. 
A planar traveling ultrasound wave with frequency $f=\SI{1}{\MHz}$ enters the simulation domain at the left boundary (the inlet), propagates in the $x_1$ direction, interacts with the particle, and leaves the domain at the right boundary (the outlet). 
The ultrasound wave entering the system at the inlet is described by the time-dependent velocity $\ws_{\mathrm{in}}(t)=(\Delta p / (\rho_0 c_{\mathrm{f}})) \sin(2\pi f t)$ and pressure $p_{\mathrm{in}}(t)=\Delta p \sin(2\pi f t)$ with the pressure amplitude $\Delta p=\SI{10}{\kilo\pascal}$ and the mass density $\rho_0=\SI{998}{\kilogram\,\metre^{-3}}$ and sound velocity $c_\mathrm{f}=\SI{1484}{\metre\,\second^{-1}}$ of the unperturbed fluid. 
This ultrasound wave has an acoustic energy density $E=\Delta p^2/(2 \rho_0 c_{\mathrm{f}}^2)=\SI{22.7}{\milli\joule\,\metre^{-3}}$. 
We choose the width of the simulation domain so that $l_1=\lambda/4$, where $\lambda=\SI{1.484}{\milli \metre}$ is the wavelength of the ultrasound. 
The orientation of the particle is described by an angle $\theta$ that is measured from the positive $x_1$ axis to the vector that runs from $S$ to the tip of the particle that is on its axis of symmetry. 
We vary the orientation from $\theta=0$, where the particle points in the direction of propagation of the ultrasound, to $\theta=\pi$, where the particle and the ultrasound wave point in opposite directions. 
The ultrasound exerts time-averaged propulsion forces $F_\parallel$ and $F_\perp$ parallel and perpendicular to the particle orientation, respectively, and a time-averaged propulsion torque $T$ on the particle. $F_\parallel$, $F_\perp$, and $T$ act on $S$. 
We denote the directions parallel and perpendicular to the particle by orientational unit vectors 
\begin{align}
\uu_\parallel&=(\cos(\theta),\sin(\theta))^\mathrm{T},\\
\uu_\perp&=(\sin(\theta),-\cos(\theta))^\mathrm{T},
\end{align}
respectively, where $\uu_\perp$ points in the negative $x_2$ direction for $\theta=0$. 
For the boundaries of the simulation domain parallel to the $x_1$ axis, we use a slip boundary condition, and for the boundary of the particle, we use a no-slip boundary condition.
The parameters of the system that are relevant for our simulations and the values assigned to these parameters are summarized in Tab.\ \ref{tab:Parameters}.
\begin{table}[htb]
\centering
\begin{ruledtabular}
\begin{tabular}{p{46mm}ccc}%
\textbf{Name} & \textbf{Symbol} & \textbf{Value}\\
\hline
Particle diameter & $\sigma$ & $2^{-1/2}\SI{}{\micro\metre}$\\
Particle height & $h$ & $\sigma$\\
Particle orientation angle & $\theta$ & $0$-$\pi$ \\
Sound frequency & $f$ & $\SI{1}{\mega\hertz}$\\
Speed of sound & $c_\mathrm{f}$ & $\SI{1484}{\metre\,\second^{-1}}$\\
Time period of sound & $\tau=1/f$ & $\SI{1}{\micro\second}$\\
Wavelength of sound & $\lambda=c_\mathrm{f}/f$ & $\SI{1.484}{\milli\metre}$\\
Temperature of fluid & $T_0$ & $\SI{293.15}{\kelvin}$ \\
Mean mass density of fluid & $\rho_0$ & $\SI{998}{\kilogram\,\metre^{-3}}$\\
Mean pressure of fluid & $p_{0}$ & $\SI{101325}{\pascal}$ \\
Initial velocity of fluid & $\ww_{0}$ & $\vec{0}\,\SI{}{\metre\,\second^{-1}}$ \\
Sound pressure amplitude & $\Delta p$ & \SI{10}{\kilo\pascal}\\
Acoustic energy density & \hspace*{-4mm}$E=\Delta p^2/(2 \rho_0 c_{\mathrm{f}}^2)$\hspace*{-1mm} & $\SI{22.7}{\milli\joule\,m^{-3}}$\\
Shear/dynamic viscosity of fluid & $\nu_{\mathrm{s}}$ & $\SI{1.002}{\milli\pascal\,\second}$ \\
Bulk/volume viscosity of fluid & $\nu_{\mathrm{b}}$ & $\SI{2.87}{\milli\pascal\,\second}$ \\
\mbox{Inlet-particle or particle-outlet} distance & $l_1$ & $\lambda/4$ \\
Inlet length & $l_2$ & $\SI{200}{\micro\metre}$\\
Mesh-cell size & $\Delta x$ & $\SI{15}{\nano \metre}$-$\SI{1}{\micro \metre}$ \\
Time-step size & $\Delta t$ & $1$-$\SI{10}{\pico \second}$\\
Simulation duration & $t_{\mathrm{max}}$ & $500\tau$ \\
\end{tabular}%
\end{ruledtabular}%
\caption{\label{tab:Parameters}Parameters that are relevant for our simulations and their values. The values of the speed of sound $c_\mathrm{f}$, mean mass density $\rho_0$, shear viscosity $\nu_\mathrm{s}$, and bulk viscosity $\nu_\mathrm{b}$ are calculated for quiescent water at standard temperature $T_0$ and standard pressure $p_0$, where the value of $\nu_\mathrm{b}$ is determined by a cubic spline interpolation of the data from Tab.\ 1 in Ref.\ \cite{HolmesPP2011}.}%
\end{table}

In our simulations, we solve a set of coupled partial differential equations consisting of the continuity equation for the mass-density field of the fluid, the compressible Navier-Stokes equations, and a linear constitutive equation for the fluid's pressure field. These direct fluid dynamics simulations are performed using the finite volume software package OpenFOAM \cite{WellerTJF1998}.

The simulations first yield the time-dependent force and torque acting on the particle as well as the flow field in the fluid. 
We calculate this force in the laboratory frame. 
The force and torque can be calculated from the stress tensor $\Sigma=\Sigma^{(p)}+\Sigma^{(v)}$ of the fluid. 
This stress tensor consists of a pressure contribution $\Sigma^{(p)}$ and a viscous contribution $\Sigma^{(v)}$ so that the force and torque acting on the particle can be written as sums $\vec{F}^{(p)}+\vec{F}^{(v)}$ and $T^{(p)}+T^{(v)}$, respectively, where $\vec{F}^{(p)}$ and $T^{(p)}$ are the contributions of $\Sigma^{(p)}$ and $\vec{F}^{(v)}$ and $T^{(v)}$ are the contributions of $\Sigma^{(v)}$. The expressions for the contributions to the force and torque are \cite{LandauL1987}
{\allowdisplaybreaks\begin{align}%
F^{(\alpha)}_{i} &= \sum^{2}_{j=1} \int_{\partial\Omega_{\mathrm{p}}} \!\!\!\!\!\!\! \Sigma^{(\alpha)}_{ij}\,\dif A_{j}, \label{eq:F}\\
T^{(\alpha)} &= \sum^{2}_{j,k,l=1} \int_{\partial\Omega_{\mathrm{p}}} \!\!\!\!\!\!\! \epsilon_{ijk}(x_j-x_{\mathrm{p},j})\Sigma^{(\alpha)}_{kl}\,\dif A_{l}
\label{eq:T}%
\end{align}}%
with $\alpha\in\{p,v\}$, the normal and outwards oriented element $\dif\vec{A}(\vec{x})=(\dif A_{1}(\vec{x}),\dif A_{2}(\vec{x}))^{\mathrm{T}}$ of the particle surface $\partial\Omega_{\mathrm{p}}$ at position $\vec{x}\in\partial\Omega_{\mathrm{p}}$, the Levi-Civita symbol $\epsilon_{ijk}$, and the center-of-mass position $\vec{x}_\mathrm{p}$ of the particle.
During a simulation, the particle is held in its position and orientation. The forces $\vec{F}^{(p)}$ and $\vec{F}^{(v)}$ and torques $T^{(p)}$ and $T^{(v)}$, therefore, correspond to a fixed particle or a particle with infinite mass density, which can be seen as limiting case for a particle with a large mass density. 

We average the forces and torques over one period $\tau=1/f=\SI{1}{\micro\second}$ for large times $t$ and extrapolate $t \to \infty$ using the extrapolation procedure that is described in Ref.\ \cite{VossW2020} to obtain the forces and torques corresponding to the stationary state. 
This results in the mean propulsion force $\vec{F}=\vec{F}_p+\vec{F}_v$ with contributions $\vec{F}_p=\langle\vec{F}^{(p)}\rangle$ and $\vec{F}_v=\langle\vec{F}^{(v)}\rangle$ as well as the mean propulsion torque $T=T_p+T_v$ with contributions $T_p=\langle T^{(p)}\rangle$ and $T_v=\langle T^{(v)}\rangle$, where $\langle\cdot\rangle$ denotes the time average.
The components $F_\mathrm{\parallel}$ and $F_\perp$ of the propulsion force that are parallel and perpendicular to the particle orientation, respectively, are calculated by the projection
\begin{align}
F_\mathrm{\parallel} &= \vec{F}\cdot\uu_\parallel , \label{eq:parallel_force} \\
F_\perp &= \vec{F}\cdot\uu_\perp . \label{eq:perpendicular_force}
\end{align}

We are also interested in the translational velocities $v_\parallel$ and $v_\perp$ and the angular velocity $\omega$ that correspond to $F_\mathrm{\parallel}$, $F_\perp$, and $T$. 
They can be calculated with the Stokes law \cite{HappelB1991}
\begin{equation}
\vec{\mathfrak{v}}=\frac{1}{\nu_\mathrm{s}}\boldsymbol{\mathrm{H}}^{-1}\,\vec{\mathfrak{F}}, \label{eq:velocity}
\end{equation}
where $\vec{\mathfrak{v}}=(v_\perp,v_\parallel,0,0,0,\omega)^{\mathrm{T}}$ is a translational-angular velocity vector, $\vec{\mathfrak{F}}=(F_\perp,F_\parallel,0,0,0,T)^{\mathrm{T}}$ a force-torque vector, $\nu_\mathrm{s}$ the shear viscosity of the fluid, and 
\begin{equation}
\boldsymbol{\mathrm{H}}=
\begin{pmatrix}
\boldsymbol{\mathrm{K}} & \boldsymbol{\mathrm{C}}^{\mathrm{T}}_{\mathrm{S}} \\
\boldsymbol{\mathrm{C}}_{\mathrm{S}} & \boldsymbol{\Omega}_{\mathrm{S}} 
\end{pmatrix}
\label{eq:H}%
\end{equation}
the hydrodynamic resistance matrix of the particle for $\theta=\pi/2$. 
In this matrix, $\boldsymbol{\mathrm{K}}$, $\boldsymbol{\mathrm{C}}_{\mathrm{S}}$, and $\boldsymbol{\Omega}_{\mathrm{S}}$ are submatrices and the subscript $\mathrm{S}$ denotes the reference point for the calculation of $\boldsymbol{\mathrm{H}}$, which is chosen here as the center of mass. 
The matrix values are calculated with the software \texttt{HydResMat} \cite{VossW2018,VossJW2019}.
Since the matrix $\boldsymbol{\mathrm{H}}$ corresponds to a three-dimensional particle, but we perform here simulations in two spatial dimensions to keep the computational effort manageable, we assign a thickness of $\sigma$ of the particle in the third dimension.  
This leads to the submatrices
\begin{align}
\boldsymbol{\mathrm{K}} &= \begin{pmatrix}
\SI{7.74}{\micro\metre} & 0 & 0\\
0 & \SI{7.48}{\micro\metre} & 0 \\
0 & 0 & \SI{7.16}{\micro\metre}
\end{pmatrix},
\label{eq:K}
\\
\boldsymbol{\mathrm{C}}_{\mathrm{S}} &= \begin{pmatrix}
0 & 0 & \SI{0.05}{\micro\metre^2}\\
0 & 0 & 0 \\
\SI{-0.11}{\micro\metre^2} & 0 & 0
\end{pmatrix},
 \\
\boldsymbol{\Omega}_{\mathrm{S}} &= \begin{pmatrix}
\SI{1.81}{\micro\metre^3} & 0 & 0\\
0 & \SI{1.69}{\micro\metre^3} & 0 \\
0 & 0 & \SI{1.73}{\micro\metre^3}
\end{pmatrix}.
\label{eq:Omega}
\end{align}
From the hydrodynamic resistance matrix $\boldsymbol{\mathrm{H}}$, we can also calculate the diffusion coefficient $\mathcal{D}=(k_\mathrm{B} T_0 / \nu_\mathrm{s}) \boldsymbol{\mathrm{H}}^{-1}$ of the particle, where $k_\mathrm{B}$ is the Boltzmann constant. The particle's rotational diffusion coefficient, corresponding to rotation in the $x_1$-$x_2$ plane, is then given by $D_\mathrm{R}=(\mathcal{D})_{33}$.  

To estimate the numerical error that is associated with our results for $v_\parallel$ and $v_\perp$, we make use of the fact that the results for the forces $\vec{F}_p$ and $\vec{F}_v$ are, due to the numerical inaccuracies of the calculations, not exactly zero for $\theta=0$ and $\theta=\pi$, although they should be for reasons of symmetry. We therefore choose the absolute values of $\vec{F}_p$ and $\vec{F}_v$ for $\theta=0$ and $\theta=\pi$ as estimates for the errors that correspond to $\vec{F}_p$ and $\vec{F}_v$. Considering the propagation of uncertainty, we then obtain from the estimated errors of $\vec{F}_p$ and $\vec{F}_v$ an estimate for the error of $\vec{F}$ for $\theta=0$ and $\theta=\pi$ each. Next, we choose the maximum of the estimated errors for both angles as an estimate for the error that corresponds to $\vec{F}$. Finally, we calculate from this error the error that is associated with $v_\parallel$ and $v_\perp$. This error is shown in Fig.\ \ref{fig:fig1} as error bars for $v_\parallel$ and $v_\perp$. 

Nondimensionalization of the equations governing our simulations leads to four dimensionless characteristic numbers: 
a Reynolds number $\mathrm{Re}_\mathrm{s}$ corresponding to the shear viscosity $\nu_\mathrm{s}$ of the fluid
\begin{equation}
\mathrm{Re}_\mathrm{s}=\frac{\rho_0 c_\mathrm{f} \sigma}{\nu_\mathrm{s}} \approx 1045,
\end{equation}
a Reynolds number $\mathrm{Re}_\mathrm{b}$ corresponding to the bulk viscosity $\nu_{\mathrm{b}}$ of the fluid
\begin{equation}
\mathrm{Re}_\mathrm{b}=\frac{\rho_0 c_\mathrm{f} \sigma}{\nu_\mathrm{b}} \approx 365,
\end{equation}
the Helmholtz number 
\begin{equation}
\mathrm{He}= \frac{2 \pi f\sigma}{c_\mathrm{f}}\approx 2.99\cdot 10^{-3},
\end{equation}
and the product $\mathrm{Ma}^2\mathrm{Eu}$ with the Mach number $\mathrm{Ma}$ and Euler number $\mathrm{Eu}$
\begin{equation}
\mathrm{Ma}^2 \mathrm{Eu}=\frac{\Delta p}{\rho_0 c_\mathrm{f}^2}\approx 4.55\cdot 10^{-6} .
\end{equation}
Note that the largest Reynolds number describing the particle motion through the fluid is close to zero:
\begin{equation}
\mathrm{Re}=\frac{\rho_0\sigma}{\nu_\mathrm{s}} \max_{\theta\in[0,\pi]}\!\Big\{\sqrt{v_\parallel^2(\theta)+v_\perp^2(\theta)}\Big\}<10^{-7}.
\end{equation}

We discretize the fluid domain as a structured mixed rectangle-triangle mesh. It has about 250,000 cells and the typical cell size $\Delta x$ ranges from $\SI{15}{\nano\metre}$ close to the particle to $\SI{1}{\micro\metre}$ far away from the particle. For the time integration, an adaptive time-step method is used. The maximum time-step size is chosen such that the Courant-Friedrichs-Lewy number 
\begin{align}
C = c_\mathrm{f} \frac{\Delta t}{\Delta x}
\end{align}
is smaller than one. This leads to a time-step size $\Delta t$ between $\SI{1}{\pico\second}$ and $\SI{10}{\pico\second}$. For getting close to the stationary state, each simulation runs for $t_\mathrm{max}= 500\tau$ or more. Because of the fine discretization in space and time compared to the large temporal and spatial domains of the system, a typical simulation needs about $36,000$ CPU core hours.

\section*{Conflicts of interest}
There are no conflicts of interest to declare.

\begin{acknowledgments}
We thank Patrick Kurzeja for helpful discussions. 
R.W.\ is funded by the Deutsche Forschungsgemeinschaft (DFG, German Research Foundation) -- WI 4170/3-1. 
The simulations for this work were performed on the computer cluster PALMA II of the University of M\"unster. 
\end{acknowledgments}

\nocite{apsrev41Control}
\bibliographystyle{apsrev4-1}
\bibliography{control,refs}
	
\end{document}